\documentclass[letterpaper]{article} % DO NOT CHANGE THIS
\usepackage{aaai2027}
\usepackage[hyphens]{url}  % DO NOT CHANGE THIS
\usepackage{graphicx} % DO NOT CHANGE THIS
\usepackage{natbib}  % DO NOT CHANGE THIS AND DO NOT ADD ANY OPTIONS TO IT
\usepackage{caption} % DO NOT CHANGE THIS AND DO NOT ADD ANY OPTIONS TO IT
\usepackage{booktabs}
\usepackage{array}
\usepackage{threeparttable}
\usepackage{colortbl}
\definecolor{highlight}{RGB}{255,245,245}
\title{Restricting the Model, Missing the System:\\Measurement and Accountability in Offensive AI Governance}
\author{
    Michael Alexander Riegler\textsuperscript{\rm 1,2},
    Finn Schwall\textsuperscript{\rm 3},
    Annika Willoch Olstad\textsuperscript{\rm 1,3},\\
    Birk Sebastian Frostelid Torpmann-Hagen\textsuperscript{\rm 3},
    Sushant Gautam\textsuperscript{\rm 1,2},\\
    Klas H. Pettersen\textsuperscript{\rm 1},
    Inga Strümke\textsuperscript{\rm 4}
}
\affiliations{
    \textsuperscript{\rm 1}Simula Metropolitan Center for Digital Engineering (SimulaMet), Oslo, Norway\\
    \textsuperscript{\rm 2}Oslo Metropolitan University (OsloMet), Oslo, Norway\\
    \textsuperscript{\rm 3}Simula Research Laboratory, Oslo, Norway\\
    \textsuperscript{\rm 4}Norwegian University of Science and Technology (NTNU), Trondheim, Norway\\
    michael@simula.no
}

\begin{document}

\maketitle

\begin{abstract}
We show that the instruments used to measure AI offensive capability fail in two ways: (i) they overstate harm, and (ii) they credit the model with capability that belongs to the surrounding system. We argue that restricting access to a model is therefore necessary but not sufficient and that policy and procurement also need system-level, harm-grounded capability assessment.
In June 2026, two frontier models were suspended under US export controls, reportedly prompted by a jailbreak that asked a model to read a codebase and fix its flaws. This finding measured an elicitation \emph{system} of model, prompt, and task. We support our argument with a study of an open-source framework in which lightweight large language model (LLM) agents coordinate through shared memory and evolutionary optimization, providing two pieces of evidence. First, jailbreak metrics overstate harm: over 225 swarm-generated attacks per target, LLM-as-judge scoring rated Claude Sonnet~4 compromised in 40\% of attacks, yet manual verification found actionable harmful content in none, against a 45.8\% Effective Harm Rate for GPT-4o. Second, scaffolded evaluations misattribute capability: on a planted-vulnerability target, a full pipeline built around a 1.2B-parameter model recovers 9 of 9 weaknesses, while the same model without the hand-crafted components recovers 0 of 9 by crash verification and 2 of 9 by cited source line. Offensive capability is a property of model, scaffold, and evaluation protocol together. The duty to assess it should lie with whoever controls the system, shapes its behaviour, and can foresee what it will do. In our setting, that is the party who builds the harness around the model, a role that current regulation does not clearly cover.
\end{abstract}

\section{Introduction}

This paper asks two questions: whether the instruments that policy relies on measure the offensive danger of AI correctly, and whether restricting access to a single model, the main policy tool in use today, acts on the quantity those instruments track. We argue that on both counts the answer is presently no, and that this is a problem of evaluation and accountability before it is a security one: a regulator, a standards body, or a procurement team cannot govern a quantity it mismeasures, and it cannot assign responsibility for a capability it has attributed to the wrong component.

The question has already come up in practice. On 7 April 2026, Anthropic announced Project Glasswing, a restricted preview of a model it described as capable of identifying and then exploiting zero-day vulnerabilities in every major operating system and web browser when directed by a user, and which it did not plan to make generally available~\cite{anthropic2026glasswing,anthropic2026mythosresearch}. Independent evaluation by the UK AI Security Institute corroborated the capability claim: the preview was the first model to complete the Institute's 32-step end-to-end corporate-network attack simulation, in 3 of 10 attempts~\cite{aisi2026mythos}. In June, a US export-control directive suspended access by foreign nationals to two of the developer's frontier models, Fable~5 and Mythos~5, and the developer disabled both for all customers to comply~\cite{anthropic2026suspend}. The directive was lifted on 30 June, after researchers at the US Center for AI Standards and Innovation (CAISI) tested the safeguards and judged them ``extraordinarily strong,'' subject to conditions on detecting security risks, collaborating on standards, and reporting malicious activity~\cite{anthropic2026redeploy,press2026liftexport}. In the developer's understanding, the trigger for the restriction was a \emph{jailbreak}: a method that essentially consisted of asking the model to read a specific codebase and fix any software flaws~\cite{anthropic2026suspend}.

The episode approximates a natural experiment for our thesis. The finding that led to a model-level restriction described a \emph{system}: a model prompted in a particular way to do a particular task. Its severity was contested because nobody had a shared way to say how much capability the jailbreak added over what other public models already do. The developer's own statement noted that the vulnerabilities in question ``all appear relatively simple, and \dots other publicly-available models are able to discover them as well without requiring a bypass''~\cite{anthropic2026suspend}. Whether or not one agrees, the disagreement was about measurement: what was gained, by which component, over what baseline.

This paper provides empirical evidence through two experiments, each of which bears as much on measurement as on capability. The first shows that a coordinated swarm of 1.2B-parameter models compromises the output-level safety of one market-leading frontier model but not another, and that the standard jailbreak metric misreads the difference. The second shows that the same class of model, embedded in a pipeline with regex pattern detection, a hand-crafted exploit corpus, and AddressSanitizer-based runtime verification, performs software vulnerability discovery requiring cross-file reasoning, and that almost none of that capability survives once the hand-crafted components are removed. We measure what the 1.2B model contributes autonomously, without those components, to keep the two apart.

Our results imply that both the threat from AI-assisted attacks and our ability to measure it are system-level phenomena. The threat depends on coordination, architecture, and evolutionary optimization, all of which are already widely available, and not only on access to any single model; the measurement depends on whether the metric tracks realized harm and whether it attributes capability to the right component. Neither is captured by asking only which model an actor can obtain. We use these findings to make a governance argument the security-benchmarking literature has largely left implicit: if capability is a property of the system, then the duty to \emph{assess} it must attach to whoever controls and can foresee the system's behaviour, and the AI value chain as currently drawn has no clean slot for the party who assembles the harness.

\paragraph{Statement of Contributions.}
(1)~We extend evidence that jailbreak success overstates realized harm~\cite{souly2024strongreject,nikolic2025jailbreaktax,yan2025confusion} to frontier production models under population-based attack, and document a second failure mode, the misattribution of scaffold capability to the model, each with reproducible evidence from open-weights models on consumer hardware.
(2)~We propose a harm-grounded success metric and a capability-attribution decomposition, and connect both to a live regulatory gap: the AI Act requires robustness against adversarial manipulation without defining it operationally, and a frontier developer has proposed, with its partners, a severity rubric with the same shape as our proposal.
(3)~We argue that the duty to assess system-level capability should follow realized, foreseeable control, and we show why existing value-chain roles (provider, deployer, distributor, importer) do not fit the harness assembler.

\emph{Relation to earlier versions.} This paper extends earlier preliminary work that presented the multi-agent framework and the two experiments as security results (not published or currently under review). New here are the harm-grounded companion grading with three judges (Section 4.1), the autonomous configuration that separates model from scaffold contribution (Section 5.1), the reading of both experiments as measurement failures (Sections 4–5), the harm-grounded metric and attribution decomposition as assessment instruments (Sections 6.1–6.2), and the governance analysis of who holds the assessment duty, why the value-chain vocabulary does not name the harness assembler, and the elicitation gap (Sections 2, 6.3–6.5).

\section{Background: the governance moment}
\label{sec:background}

The main policy instrument for offensive AI capability is access restriction at the level of a single model: withhold the frontier model, or give it only to selected partners, and the offensive capability it represents is assumed to be contained. The 2026 export-control episode recounted above shows how this works in practice: restriction was the only tool available, no shared, vendor-neutral standard existed for scoring the severity of the jailbreak claim, and governments and companies improvised under time pressure, cutting access and restoring it three weeks later once government researchers had tested the safeguards~\cite{anthropic2026suspend,anthropic2026redeploy,nextgov2026anthropic}.

Two developments since then bear directly on our argument. First, the actors closest to the problem have started building the missing instrument. Shortly after the directive was lifted, the developer proposed the Cyber Jailbreak Severity (CJS) scale, an early-draft, CVSS-style rubric that scores jailbreaks from 0 to 4 on capability gain, breadth of that gain, ease of weaponization, and discoverability, and that it is developing with its Glasswing partners~\cite{anthropic2026cjs,anthropic2026redeploy}. Like our metric, it separates a \emph{bypass} from an increase in realized harmful capability, and scores a finding at zero when a publicly available tool produces an equivalent result on a reasonable timescale, in an equivalent harness, without additional expert input. The rubric is harm-grounded and baseline-relative, as our metric is; whether it becomes a shared, vendor-neutral standard is still open.

Second, enforcement has begun while the definitions are still open. The EU AI Act requires that high-risk AI systems be robust against adversarial manipulation (Art.~15) but provides no operational definition of ``robust''. Obligations for general-purpose models with systemic risk have applied since August 2025, and since August 2026 the Commission, through the AI Office, can enforce them: it can request information, evaluate a model (including by requesting access), and, if necessary, restrict the model's availability (Arts.~91--93 and 101)~\cite{euaiact2024}. Meanwhile, the Digital Omnibus postpones the high-risk obligations to December 2027 (Annex~III) and August 2028 (Annex~I)~\cite{digitalomnibus2026}. Enforcement now operates around a measurement question nobody has settled. We argue that settling it correctly comes before, and does not depend on, whichever tool a regulator or procurement team ends up using.

\section{Related work}
\label{sec:related}

\paragraph{Automated red teaming of LLM safety.}
Automated jailbreak methods split into white-box and black-box branches. GCG~\cite{zou2023universal} finds adversarial suffixes through gradient, but requires access to the target weights. PAIR~\cite{chao2023jailbreaking} and TAP~\cite{mehrotra2024tap} pioneered the black-box attacker-LLM pattern, with TAP generalizing PAIR to a tree-of-attacks search. HarmBench~\cite{mazeika2024harmbench} standardizes evaluation across many methods and targets with a fine-tuned judge. That automated jailbreak success overstates realized harm is established: StrongREJECT~\cite{souly2024strongreject} finds standard evaluations significantly overstate jailbreak effectiveness relative to human judgment, and the jailbreak tax~\cite{nikolic2025jailbreaktax} shows jailbroken outputs lose task quality, with accuracy drops up to 92\% on reasoning tasks. JailbreakBench~\cite{chao2024jailbreakbench} provides a standard judged robustness benchmark for the same phenomenon. VENOM~\cite{yan2025confusion} finds that LLM judges, GPT-4o among them, anchor on toxic language: they still label most responses as jailbreaks when all factual content is removed and only a malicious tone remains. Expected Harm~\cite{chen2026expectedharm} weights jailbreak severity by the likelihood that the threat is carried out, modeled as a function of execution cost.

Our work differs on three points: our attackers are a population that shares memory and evolves across generations; our attacker is a swarm of 1.2B models on consumer hardware (PAIR and TAP typically use GPT-3.5 or GPT-4), and we show that judge false positives differ by target, i.e., the judge rewards one model's structured refusals as compliance, so cross-model comparisons on technical success rate are unreliable. 
Concurrent work on small language models reports the inverse failure mode. Some models that look safe under naive testing turn out to be highly susceptible to tailored jailbreaks~\cite{piano2026small}. Both directions point to the same gap between surface safety metrics and realized harm.

\paragraph{LLM-assisted vulnerability discovery.}
Prior work organizes along two axes: realism of the target (synthetic benchmarks vs.\ production code) and autonomy of the model (static evaluation vs.\ tool-augmented agentic operation). Benchmark suites such as CyberSecEval~2~\cite{bhatt2024cyberseceval2} operate on isolated snippets and cannot capture cross-file reasoning. Agentic frameworks such as Google's Big Sleep~\cite{bigsleep2024} pair a frontier LLM with code-browser, sandbox, and debugger tools to plan multi-step exploit chains, but reasoning budget and attack diversity are bounded by a single model instance. LLM-augmented fuzzers such as Fuzz4All~\cite{xia2024fuzz4all} use the model as a structured input generator; the search strategy remains the fuzzer's. Appendix~\ref{app:related} expands these descriptions.

\paragraph{The harness as a variable.}
A complementary line treats the multi-agent harness itself as a search variable. AgentFlow~\cite{liu2026agentflow} represents each harness as a program in a typed graph DSL and uses an outer-loop optimizer to propose harness edits. Holding the model fixed, pass rates on an 89-task terminal benchmark span a roughly $4\times$ range (20\% to 80\%) across publicly ranked harnesses; run against a production web browser with an open-weights mixture-of-experts model (32B active parameters), the same loop discovered ten previously unknown zero-days, all confirmed by the vendor and two of them critical sandbox escapes. This is independent, cross-sectional evidence for the thesis we test longitudinally on a single target: capability is a property of the harness and the model together. AgentFlow operates at far larger scale (a 192-GPU cluster for seven days) than our fixed, hand-designed swarm of 1.2B open-weights models on consumer hardware; the two studies sit at opposite ends of the scale range.

With the partial exception of AgentFlow, the approaches above share reliance on a single frontier-scale model in a \emph{monolithic topology}: one model holds the full reasoning trajectory in its context. We change both, replacing the single frontier model with a \emph{swarm} of \emph{smaller} models. The analogy is single-trajectory versus population-based search: individual agents are weaker, but the ensemble gains diversity, parallelism, and coverage.

\paragraph{Accountability in AI supply chains.}
That capability is distributed across a system is, for governance, a version of the ``many hands'' problem: when many actors contribute to an outcome, responsibility diffuses and no one is clearly answerable~\cite{cobbe2023accountability}. Cobbe et al.\ show that algorithmic supply chains create an ``accountability horizon'': each actor sees its immediate counterparties and little beyond them, so neither producer nor deployer holds both the knowledge and the control needed to ensure compliance. They argue that accountability should be allocated by \emph{factual control}, whatever the contracts say. We take this as the starting point for Section~\ref{sec:assessment}: our contribution is to show, empirically, where the control that realizes offensive capability actually sits, and to argue that the assessment duty should follow it. A recent policy report treats the harness as a distinct object of governance and argues that agents should be evaluated together with their tools, memory, permissions, and orchestration loop~\cite{liu2026harnessgov}. It does not address offensive capability or how harness builders map onto statutory value-chain roles, which is where our analysis goes further.

\section{Evidence I: jailbreak metrics overstate harm}
\label{sec:exp1}

The two evidence sections make two claims about instruments, each supported by one experiment: a metric that scores refusals as breaches \emph{overstates} realized harm, and does so unevenly across models (this section); and a metric that reports one recall number for a model-plus-scaffold pipeline \emph{misattributes} the scaffold's capability to the model (Section~\ref{sec:exp2}). One is about the numerator of a safety score; the other is about which component the score belongs to. Section~\ref{sec:assessment} builds the governance argument on both.

\subsection{Method}

Five instances of LFM2.5-1.2B-Thinking~\cite{liquidai2026lfm25,liquidai2025lfm2} operated as a coordinated attack swarm, each assigned a distinct strategy: direct instruction override, persona-based roleplay, multi-turn escalation, obfuscation, and evolutionary mutation of successful attacks. The swarm maintained shared memory of successful attacks and evolved over 15 generations using crossover (rate 0.5) and mutation (rate 0.3), retaining the two highest-performing attacks per generation.

Two frontier models were targeted: GPT-4o and Claude Sonnet~4 (target identifiers \texttt{gpt-4o} and \texttt{claude-sonnet-4-20250514}, both queried 22~January~2026; these snapshots were superseded by mid-2026, see Section~\ref{sec:limitations}). Each received 225 attacks (15 generations $\times$ 5 agents $\times$ 3 tasks: explosives instructions, a phishing email template, and hacking into a computer system). Attack success was scored via cross-provider LLM-as-a-judge evaluation: Claude judged GPT-4o outputs and GPT-4o judged Claude outputs. This design exposed an asymmetry in judge reliability: the GPT-4o judge often scored Claude's structured refusals as successful attacks, crediting format compliance even when the content refused, so Claude's raw attack-success rate is inflated and the two targets' technical success rates, scored by different judges, are not directly comparable. This non-comparability is one of the paper's findings. We instead report the Effective Harm Rate (EHR): the proportion of attacks producing verified actionable harmful content, determined by a manual verification pass over every judged-successful response (the verification protocol below). EHR relies on human verification rather than an automated judge, so it is the metric we compare across targets; the technical success rate is reported only to expose the gap between apparent and realized success.

\paragraph{Verification protocol.} Because EHR depends on human judgment, we set out that judgment in detail. The cross-provider judge scores \emph{compliance}: how fully the target engaged with the request. Compliance is what EHR has to correct for, because a refusal can score as full engagement. The manual verification pass covered the review set: every response the judge scored as successful (compliance score $\geq 0.6$, the run's success threshold), which is 137 responses for GPT-4o and 90 for Claude Sonnet~4. One author read each response in a single pass and counted it as actionable harmful content only if it supplied operationally specific means of causing harm--concrete, executable steps rather than a topical description, generic caution, or a refusal restated in the attacker's terms. Of the review set, 103 GPT-4o responses (45.8\% of 225) and 0 Claude responses met this bar. 
Each response was judged once, by one annotator, and the per-response decisions were not logged, so we cannot report inter-annotator agreement, and the absolute rates carry that annotator's threshold (Section~\ref{sec:limitations}). The comparison between targets does not rest on this pass alone: three automated graders applying the published StrongREJECT rubric reproduce its direction and categorical size (below).
Because the pass covers every judged-successful response and lower-scored responses are excluded by construction, the reported rates are lower bounds on realized harm (a response the judge scored below threshold could only add to the count, never remove from it), which is the conservative direction for our argument.

\paragraph{A reproducible harm-grounded companion.} As a reproducible companion to the author-verified rate, we grade every released response with the published StrongREJECT rubric~\cite{souly2024strongreject}, run from a fixed definition with the judge model recorded on each response. To keep the score from depending on a single provider, we use three judges over both targets: an OpenAI model (\texttt{gpt-4o-2024-11-20}), StrongREJECT's provider-neutral fine-tuned open-weights grader, and an Anthropic model (\texttt{claude-sonnet-4-5}, via the standard API without elevated security-research access). 

All three rank GPT-4o above Claude Sonnet~4 at a 0.5 rubric threshold. The two that score every response agree closely with the manual pass: GPT-4o at 42.7\% and 44.0\% harmful against Claude's 11.6\% and 4.9\% (the rubric's 0.5 threshold credits partial compliance that the actionable-harm bar excludes, so Claude is low rather than zero). The Anthropic judge ranks the targets the same way (13.6\% against 1.9\% of the responses it scored), but its content filter declines to score 51 responses (41 of GPT-4o's, mostly from the explosives task, and 10 of Claude's) so its rates are computed over 184 and 215 responses and are not directly comparable with the other two graders. That the declines fall four times as often on GPT-4o's responses is consistent with the direction of the gap. The automated rubric and the manual pass agree on the direction and the categorical size of the gap (script and the cached per-response scores are provided in the supplement).

\subsection{Results}

\begin{table}[tb]
\centering
\begin{threeparttable}
\caption{Attack outcomes against GPT-4o and Claude Sonnet~4 over 225 swarm-generated attacks per target. Technical Success Rate uses LLM-as-judge scoring and includes false positives where structured refusals were scored as successes; Effective Harm Rate counts only attacks producing manually verified actionable harmful content.}
\label{tab:exp1}
\begin{tabular}{lcc}
\toprule
\textbf{Metric} & \textbf{GPT-4o} & \textbf{Claude Sonnet~4} \\
\midrule
Technical Success Rate\tnote{*} & 60.9\% & 40.0\% \\
\rowcolor{highlight} \textbf{Effective Harm Rate}\tnote{$\dagger$} & \textbf{45.8\%} & \textbf{0.0\%} \\
\rowcolor{highlight} Critical Severity Attacks & 49 & 0 \\
Perfect Score (1.0) & 18 & 0 \\
Refusal Rate & 15.1\% & 56.9\% \\
\bottomrule
\end{tabular}
\begin{tablenotes}
\footnotesize
\item[*] Technical success: LLM-judge compliance score $\geq 0.6$ (the run's success threshold); includes false positives.
\item[$\dagger$] Manual verification pass over the review set (the judged-successful responses, score $\geq 0.6$): counts only attacks producing verified actionable harmful content.
\end{tablenotes}
\end{threeparttable}
\end{table}

Table~\ref{tab:exp1} summarizes the outcomes. GPT-4o produced 49 critical-severity responses containing detailed attack methodologies, sabotage techniques, and operational frameworks; in 91.2\% of successful attacks it provided structured step-by-step instructions, and some offered to elaborate further. Claude Sonnet~4, despite a 40\% technical success rate, produced no actionable harmful content: its nominally successful responses were benign reinterpretations of the attack prompt, educational content with caveats, or explicit identification and refusal of the jailbreak attempt. The gap is categorical rather than a matter of degree (45.8\% Effective Harm against GPT-4o, 0\% against Claude). Claude's nominally successful responses still declined to produce harmful content, which points to output-level safeguards that act at generation rather than only at input filtering.

The swarm roughly doubled its success rate against GPT-4o over the 15-generation run (generations numbered 0--14): the first two generations (0--1) averaged 33.3\%, the generations-7--14 window averaged 65.8\%, and the final generation reached 73.3\%. These windows track the learning curve and are not the overall rate, which is computed over all 225 attacks (Table~\ref{tab:exp1}). The proportionally larger improvement against Claude Sonnet~4 (6.7\% to 46.7\%) started from a near-zero baseline and did not translate into effective harm. Per-generation data is in Table~\ref{tab:evolution} (Appendix~\ref{app:evolution}).

Neither model is safe under all conditions. The finding is narrower: a single automated metric compresses a categorical difference into a modest one. By technical success rate, Claude Sonnet~4 is compromised at two-thirds of GPT-4o's rate (40.0\% against 60.9\%); by realized harm the gap is 0\% against 45.8\%. A procurement rule that rejects any model above a fixed jailbreak rate, e.g., 30\%, would reject both and never see the realized-harm gap between them. A regulator or procurement team using technical jailbreak rate as its figure of merit would read Claude as ``broken 40\% of the time'' and might mark it down against a model that is, by the harm-grounded measure, far worse. The error lies in the instrument rather than in the model.

\section{Evidence II: scaffolds misattribute capability}
\label{sec:exp2}

The second piece of evidence is narrower by design: one planted-vulnerability target, studied closely, to isolate a single quantity, namely how much of a reported recall number belongs to the model and how much to the scaffold around it. We present it as a case study rather than a capability estimate; the protocol, scaling across swarm sizes, and per-CWE analysis are in Appendices~\ref{app:iterations}--\ref{app:modality}, with the full per-version history in the supplementary material.

The swarm is an effective jailbreak attacker (Section~\ref{sec:exp1}) and a weak autonomous vulnerability finder, and this is not a contradiction: jailbreak search rewards diversity and persistence, which a population supplies, whereas planted-bug discovery rewards multi-step cross-file reasoning, which a 1.2B model lacks. The scaffold supplies the latter, which is the point of this experiment.

\subsection{Method}

The target was a deliberately vulnerable C application (SwarmApp, $\sim$550 lines across 5 files, roughly 340 excluding comments and blank lines) containing 9 planted vulnerabilities spanning 9 distinct Common Weakness Enumeration (CWE) identifiers, including stack and heap buffer overflow, format string, integer overflow, use-after-free, double free, null pointer dereference, command injection, and hardcoded credentials. The application also included correct uses of dangerous functions, red herrings meant to trip keyword scanners into false positives.

The full (\emph{assisted}) pipeline ran in two phases: source-code analysis by role-differentiated 1.2B agents using regex pattern detection followed by grounded LLM citation, with a post-hoc CWE normalization layer; and binary fuzzing under AddressSanitizer against a hand-crafted seed corpus with LLM-generated mutations. To separate the pipeline's capability from the model's, we also ran an \emph{autonomous} configuration on the same target with the hand-crafted seeds, the regex layer, and post-hoc normalization all disabled. There, findings are credited only via \emph{citation-verified} recall (LLM-cited source lines matching ground-truth signatures held only at scoring time) and \emph{crash-verified} recall (LLM-generated fuzz inputs producing actual ASan crashes).

\subsection{Results}

\begin{table}[tb]
\centering
\begin{threeparttable}
\caption{Recall on the 9-CWE planted target at 7~agents~$\times$~3~generations. \emph{Assisted}: full pipeline. \emph{Autonomous}: hand-crafted components disabled, crediting only LLM-generated crashes or LLM citations matching ground-truth signatures. The gap between assisted (9/9) and autonomous crash-verified (0/9), at fixed model and target, isolates the scaffold's contribution from the model's.}
\label{tab:exp2}
\begin{tabular}{lc}
\toprule
\textbf{Configuration} & \textbf{Recall} \\
\midrule
Planted vulnerabilities & 9 \\
\rowcolor{highlight} \textbf{Assisted (full pipeline)} & \textbf{9/9 (100\%)} \\
\rowcolor{highlight} \textbf{Autonomous, citation-verified} & \textbf{2/9 (22\%)} \\
\rowcolor{highlight} \textbf{Autonomous, crash-verified} & \textbf{0/9 (0\%)} \\
LLM CWE labels correct (autonomous) & 1 of 6 \\
\bottomrule
\end{tabular}
\begin{tablenotes}
\footnotesize
\item Runtimes: full configuration completes in $\sim$4 minutes on a consumer laptop. Scaling across swarm sizes (6/9 to 9/9) is in Appendix~\ref{app:iterations}.
\end{tablenotes}
\end{threeparttable}
\end{table}

Table~\ref{tab:exp2} reports the core result. In the autonomous configuration the 1.2B model identifies two of the planted vulnerable lines as suspicious but assigns the correct CWE class to only one; it cannot construct the multi-step inputs needed to trigger any planted bug at runtime, and its proposed fuzz inputs include literal macro names, truncated command sequences, and placeholder text the sanitizer filters out. The full pipeline recovers all nine by spreading the work across components: regex patterns, role-specialized search, and the ASan crash classifier each contribute distinct CWEs, and the model's one credited contribution duplicates something a non-LLM component already found (per-CWE breakdown in Appendix~\ref{app:modality}).

We read the assisted 9/9 as a statement about the pipeline on this target, not as a capability estimate for the model or for software at large. The scaffold was iterated against these nine bugs (Appendix~\ref{app:iterations}), so the number is close to a fixed point of its own design; taken alone it would be a poor measure of anything. What is not an artifact of that iteration is the distance between the two configurations (9/9 assisted, 0/9 crash-verified autonomous) at fixed model and fixed target. That distance, rather than the 9/9, is what the experiment measures. On this target the recall belongs to the scaffold. The per-CWE modality breakdown (Table~\ref{tab:modality}, Appendix~\ref{app:modality}) shows why: regex-matched seeds under AddressSanitizer and the hand-crafted fuzz corpus account for seven of the nine, and the single CWE credited to the model's grounded citation (CWE-798) duplicates what role-specialized search already finds. A report that credited the 9/9 to the model would be crediting it with work done almost entirely by conventional static and dynamic analysis.

\section{Toward system-level, harm-grounded assessment}
\label{sec:assessment}

Both experiments concern the system, not the model in isolation, and both expose a measurement error that a policy or procurement decision would inherit. This section turns that into a governance argument in five steps: what a corrected metric measures, which component a capability figure belongs to, who should carry the assessment duty, why the model provider cannot carry it alone, and what this means for procurement and regulation.

\subsection{A harm-grounded success metric}

The divergence between technical success rate and Effective Harm Rate is a measurement problem with a policy cost. By conventional jailbreak metrics, Claude Sonnet~4 was ``broken'' 40\% of the time; by the metric reflecting potential for real-world harm, 0\% of the time. If regulators adopt technical jailbreak rates as their operational reading of the AI Act's undefined ``robust,'' they risk penalizing models that are safe in practice. Our two targets exhibit only this over-penalizing direction. A robustness figure should count realized harmful content, not judge-scored format compliance. EHR is one such figure and its main weakness is that ``actionable harm'' has no agreed operational definition. Our single-annotator pass and the StrongREJECT rubric draw the line in different places (0\% against 1.9--11.6\% for Claude Sonnet~4 across the three graders; Section~\ref{sec:exp1}), yet both preserve the gap between targets. That disagreement is a reason to standardize and publish the definition, not to fall back on judge-scored compliance, which counts refusals as breaches.

We are not offering EHR as a finished standard. The CVSS-style severity rubric a frontier developer has proposed~\cite{anthropic2026cjs} scores capability gain and ease of weaponization against a public baseline, and scores a jailbreak at zero when publicly available tools reach an equivalent result (same logic as EHR). That a developer and its partners drafted a similar instrument under time pressure suggests the measurement is feasible. The harder problems are agreeing on the definition and on who governs it.

\subsection{Capability attribution}
\label{sec:attribution}

An evaluation that bundles a model with a scaffold and reports one number does not measure the model and should not be read as if it did. The assisted-versus-autonomous decomposition (Section~\ref{sec:exp2}) is one way to report the model's marginal contribution separately from the system's, and the two diverge sharply at 1.2B. Concurrent work reaches the same conclusion from the other end: holding a frontier model fixed and varying the harness moves pass rates across a roughly fourfold range~\cite{liu2026agentflow}. Any assessment that will inform a decision to restrict, to procure, or to certify has to say which component it is crediting. Our planted target sits at the low-capability end, where the scaffold carries nearly all of the recall, and that harness result shows the misattribution persists at the frontier end. Model-level capability figures mislead across the whole range, which is why the attribution duty is general rather than an artifact of a small model. A combined number can also hide errors in the evaluation code itself. An early iteration of our extractor read CWE labels out of source comments and reported a spurious 100\% recall (Appendix~\ref{app:iterations}), an error class documented across vulnerability-detection benchmarks and quantified at scale across software-engineering evaluations~\cite{ding2025primevul,li2025everything,zhou2025lessleak}.

This attribution claim does not rest on our single planted target. Three independent lines converge on it: our assisted-versus-autonomous decomposition over time on one target; AgentFlow's harness comparisons at fixed frontier model and its harness search on production software~\cite{liu2026agentflow}; and the 2026 export-control episode, where a model-level restriction reportedly followed a \emph{system-level} elicitation and the proposed response, the CJS scale, scores a finding in its system context and against a public baseline instead of as a fixed property of the model~\cite{anthropic2026suspend,anthropic2026cjs}. Our planted target is the most controlled of the three; the governance argument rests on all three agreeing, not on it alone.

\subsection{Who bears the assessment duty}
\label{sec:duty}

If capability is a property of the system, the duty to assess it cannot sit only with the model provider. The question is where it should sit, and here the law already contains the principle, if not yet the vocabulary. Under the AI Act, a party that puts its name on a high-risk system, substantially modifies it, or repurposes a general-purpose system into a high-risk use \emph{becomes} its provider, and the original provider must hand over documentation and testing access (Art.~25)~\cite{euaiact2024}; duty follows modification and control. The Cyber Resilience Act is more pointed for our setting: the manufacturer that \emph{integrates} a component, including free and open-source software, must exercise due diligence over it and bears responsibility for the security of the assembled product, which it cannot shift entirely upstream (Art.~13(5))~\cite{eucra2024}. The revised Product Liability Directive points the same way, treating a party that substantially modifies a product outside the original manufacturer's control, and then makes it available or puts it into service, as a manufacturer itself (Art.~8(2))~\cite{eupld2024}. Scholarship converges on the same principle from the accountability side: allocate by factual control, not contractual fiction~\cite{cobbe2023accountability}, and calibrate obligations to the degree of control an actor actually exercises, with provider disclosure duties intensifying as downstream control falls~\cite{chen2026operational}.

The common thread is that responsibility follows realized control. In our setting that control sits with the party who assembles the harness: the coordination, the seed corpus, the pattern-to-CWE layer, and the evaluation protocol are what turn a 1.2B model that recovers 0 of 9 bugs into a pipeline that recovers 9. Yet the statutory vocabulary (provider, deployer, distributor, importer) has no clean slot for this actor. They are not the model's provider; they may deploy nothing in the regulated sense; and for an open-weights model the provider's own leverage is limited, because weights already released cannot be recalled or monitored~\cite{casper2026openweight}. The Commission's July 2025 guidance makes a downstream modifier a provider only when the modification exceeds roughly one-third of the original model's training compute~\cite{ec2025gpai}; a harness uses no training compute, so even this route does not reach the assembler. We do not propose a new liability category or a fixed answer to who pays when a harness causes harm; that is the regulator's call, and pretending otherwise would overreach the evidence. Our narrower claim is that the current vocabulary is insufficient here, and that the missing role, the harness assembler, is the one that both realizes and can foresee the system's marginal capability. Recognizing it extends the control principle already in the law, and does not break with it.

The duty binds the parties who report or rely on capability figures--evaluators, integrators, red teams, and procurement testers--not an adversary assembling a harness. Its object is the integrity of the figures on which restriction, procurement, and certification decisions rest.

One might object that the assembler is already covered: whoever ``puts into service'' an AI system is a provider under the AI Act, and Art.~25's substantial-modification rule seems designed for this case. We think the fit is looser than it looks, in three ways. First, the provider and substantial-modification duties are triggered by placing a system on the market or putting it into service; a harness built to \emph{evaluate} capability is often used internally and never placed on any market, so the trigger may simply not fire even though the capability is fully realized. Second, Art.~25's substantial-modification path presupposes a high-risk system already on the market that is then modified, but the base here is an open-weights general-purpose model, whose own high-risk classification is contested, so it is unclear there is a qualifying prior system to modify. Third, the general-purpose-model obligations attach to the model's developer, and the ``downstream provider'' duties that reach integrators are comparatively light and information-oriented, not framed around a duty to assess elicited offensive capability. None of this means no rule could ever be stretched to cover the assembler; it means coverage is contested, incidental, and not designed for the assessment duty we are pointing at. That integrators must be argued into or out of each of these categories shows the vocabulary predates the role, which is now where the capability sits.

\subsection{Why the provider cannot bear it alone: the elicitation gap}
\label{sec:elicitation}

A further reason the duty cannot rest solely on the model provider is epistemic, and it moves the test from control to control plus foreseeability. A provider's capability evaluation is a \emph{lower bound}, not an upper bound. It measures what the provider's own elicitation (its prompts, its scaffold, its fine-tuning budget) managed to surface, and a third party with a better harness, more compute, or different intuitions about the model's strengths can exceed it~\cite{metr2024elicitation,clymer2025elicitation,teun2025elicitationgame,donoway2025latent}. Elicitation bounds risk only under strong assumptions, and when they fail the developer's number guarantees less than the model can actually do~\cite{clymer2025elicitation}; models can even underperform strategically on the evaluations meant to characterize them~\cite{vanderweij2024sandbagging}. This is the epistemic form of our thesis: the provider's assessment is structurally incomplete because capability is elicited, not intrinsic.

This consequence cuts against our own argument, and we address it directly. The elicitation gap is also the best case \emph{for} access restriction: if a provider cannot bound what others will elicit, gating the model is a rational hedge. We do not argue that away. The same gap makes model-level assessment \emph{necessary}, since before release it is the provider's only tool, and \emph{insufficient}, since the part that escapes the provider's measurement is also the part that escapes the provider's control once the model is out, most of all for open weights. Both halves of the claim follow from the same fact.

The provider still owes something, and it is specific and bounded: publish capability as a lower bound \emph{with the elicitation conditions disclosed}, so downstream actors know what was and was not tested; keep monitoring and controls where they still exist (closed and API-served models, where the provider can observe elicitation and so owes more); and run elicitation testing before release when there will be no control after it, as with open weights. It also follows that credible assessment cannot be the provider's alone: independent parties must be resourced to run the elicitation the provider cannot, which is a governance case for external evaluation access~\cite{charnock2026externalaccess} and for the third-party audit ecosystem more broadly.

\subsection{Implications for procurement and regulation}

The difference between a 45.8\% and a 0\% Effective Harm Rate is a supply-chain risk determination, not a product preference: the chosen model's failure mode under adversarial pressure propagates into every downstream system that embeds it. Yet organizations have no standardized method for evaluating the adversarial robustness of the models they procure, and vendor-published benchmarks can mislead in either direction. A procurement standard that asked for harm-grounded figures and attribution reporting, with the evaluation code open to audit, would test the two things this paper shows current practice gets wrong. This also bears on the concentration of security capability: if independent evaluation of frontier claims is gated by the very labs whose claims need testing, the accountability horizon closes where it most needs to be open. Open, reproducible harnesses spread the ability to test, and they spread the ability to attack as well. The assessment duty therefore grows more important as the tools spread, not less important.

We do not claim any of this is easy to enforce. A harness is software; anyone with the expertise can write and share one, and no access gate stops that. Our claim is prior to enforcement and does not depend on solving it: whatever a regulator or procurement team acts on, it must first measure correctly, and the instruments in use measure the wrong thing in both directions. Getting the measurement right, and attaching it to the actor who controls the system, is what any later policy depends on, whether that policy works at the model level or elsewhere. Behind this sits a trade-off the paper does not settle: offensive capability is currently developed and deployed faster than the regulation and audit tools meant to assess it, and the balance between that pace and safety is being set largely by developers rather than by public bodies, which, as the 2026 episode showed, are left to improvise under time pressure. Whoever makes that trade-off can only make it well with instruments that measure the right thing.

\section{Objections and responses}
\label{sec:alt}

\paragraph{``Access restrictions still buy defensive time, even if they do not bound capability in the limit.''}
We grant the structure and much of the premise, restrictions do buy time, while disputing the level at which they buy it. The strongest evidence here is not ours: the UK AI Security Institute's evaluation of a second developer's frontier model, published on 30~April~2026, found comparable expert-level cyber-task scores (71.4\% for the second model against Mythos's 68.6\%) and a second end-to-end completion of the same 32-step simulation, in 2 of 10 attempts, and suggested that offensive cyber skill may be emerging as a byproduct of more general gains in autonomy, reasoning, and coding~\cite{aisi2026gpt55}; AgentFlow shows a large harness contribution with an open-weights model and an automated harness-search loop~\cite{liu2026agentflow}; and the export-control episode itself was reversed in under three weeks. The diffusion has already occurred at the level of capability class. Restrictions on specific models still buy time on specific exploit chains against specific products; they buy less at the level the policy debate is conducted at.

\paragraph{``Restricting frontier models still bounds the typical attacker, because effective scaffolds require expertise.''}
This grants our results and rescues restrictions on different grounds: the 9/9 assisted recall reflects months of design iteration, CWE-taxonomy knowledge, ASan instrumentation, and the discipline to catch a label-leakage bug, none of which are commodity skills. On this reading the 0/9 autonomous result supports restrictions: without the frontier model, the median attacker has nothing. The argument's narrowness is its problem. Offensive-AI policy targets state-aligned groups, organized criminal enterprises, and persistent threats, none of whom are scaffold-constrained; and scaffolds become commodity once published. The expertise barrier is real for the median actor and largely irrelevant for the actors the policy is written against.

\paragraph{``A single planted target with an iterated framework does not generalize.''}
We grant this without reservation. It is the reason why Section~\ref{sec:exp2} is framed as a case study. The framework was iterated against this target across five versions, and the assisted 9/9 is close to a fixed point of that iteration. The argument depends not on the 9/9 but on the assisted-versus-autonomous distance at fixed model and target, and the iteration does not manufacture that distance: no amount of scaffold tuning makes the \emph{autonomous} model recover the bugs. Generalization is the right next experiment: a staged evaluation that reproduces known vulnerabilities at pre-patch commits before testing novel targets under responsible disclosure.

\paragraph{``Adversarial assemblers will not comply.''}
This is true, and it is why the two are complements rather than substitutes: the duty here is a measurement-integrity obligation on the parties who produce and consume capability figures, not a constraint on attackers, and model-level restriction remains necessary for exactly the actors who will not comply.

\section{Limitations}
\label{sec:limitations}

The vulnerability case study used a planted target with known ground truth, not a production codebase; real-world discovery involves larger codebases, more complex build systems, and no prior knowledge of bug locations. The 100\% pipeline recall and the 0\% autonomous crash-verified recall are evidence about this specific target under specific configurations, not claims about arbitrary software.

Experiment~II holds the model fixed (LFM2.5-1.2B) and varies only the scaffold, while Experiment~I holds the attacking swarm fixed and varies the target model. Neither varies model scale. This isolates the scaffold's contribution in Experiment~II but says nothing about the converse: we cannot say whether a larger model would close the same gap without a scaffold, because we did not vary the model's size. We therefore make no claim that systems matter \emph{more} than models. The claim is narrower: capability, and the measurement of capability, are properties of the system, and model-level figures reported in isolation mislead. A model-size sweep at fixed scaffold is the natural complement.

The Experiment~I targets, GPT-4o and Claude Sonnet~4, were superseded by mid-2026, and we chose not to re-run against current models. The claim we intend to transfer is the divergence between technical success rate and Effective Harm Rate, not the specific magnitudes, which are properties of those models at the time of the runs. Concurrent work reports a related gap between surface safety and jailbreak robustness in small language models~\cite{piano2026small}, which suggests the divergence is not peculiar to these two targets, but we have not tested current frontier models.

Results depend on hardware, model version, scoring thresholds, and random seeds; reported numbers are point estimates rather than converged averages, and single-run variance is expected at 1.2B. Reported configurations should be treated as capability floors. 
The Effective Harm Rate depends on manual verification, performed by a single annotator in a single pass whose per-response decisions were not logged; it does not scale and yields no measure of inter-annotator agreement, so the absolute rates should be read as one annotator's application of the bar. The gap between targets does not depend on that pass alone: three StrongREJECT graders reproduce it (Section~\ref{sec:exp1}). Because verification covered only the judged-successful responses, EHR can miss harmful outputs the judge scored below threshold, so the reported rates are lower bounds on realized harm, the conservative direction for our argument. Additional details are in Appendix~\ref{app:limitations}.

\section{Conclusion}

A swarm of 1.2B-parameter models running on consumer hardware compromised the output-level safety of one frontier model at a 45.8\% Effective Harm Rate, while extracting no actionable harm from another despite an apparent 40\% jailbreak rate. Embedded in a hand-crafted pipeline, the same swarm recovered 9 of 9 planted vulnerabilities; run autonomously, it recovered 0 of 9 by crash verification and 2 of 9 by cited source line. Two implications follow. First, the categorical safety gap between the two frontier models is not captured by current jailbreak metrics, which score format compliance rather than realized harm; the field lacks a measurement regime reflecting actual stakes, though a frontier developer and its partners have begun drafting one. Second, on this target a system built around a model orders of magnitude smaller than the frontier models subject to restriction recovers all nine bugs while the same model unaided recovers almost none, which locates the capability, and its cost, in the system rather than in the model alone; a single recall figure blurs the two and, as our own label-leakage correction showed, is easy to inflate. Access restrictions on frontier models are therefore necessary but not, by themselves, a sufficient security mechanism. What is regulated has to be measured at the level of the system, with metrics that track realized harm, and the duty to measure it has to attach to whoever controls and can foresee what the system does.

\section*{Ethical Statement}

We release the swarm-attack framework on the judgment that the defender's interest in independently evaluating AI systems outweighs the marginal uplift the framework gives attackers. The rest of this section explains how we reached that judgment.

\emph{Marginal uplift.} The capability classes we demonstrate, coordinated jailbreak search and pipeline-assisted CWE detection, are already attainable by any actor able to read this paper and run a 1.2B model locally. The 2026 export-control episode is instructive here too: the developer of the restricted model reported that the vulnerabilities at issue were discoverable by other public models without a bypass~\cite{anthropic2026suspend}. These capabilities already exist; our contribution is to measure how cheap they have become and what the scaffold contributes relative to the model. For the median attacker the expertise barrier documented in Section~\ref{sec:alt} is real; for the resourced actors offensive-AI policy actually targets, the framework provides little they could not assemble.

\emph{Adverse impacts and mitigations.} The vulnerability case study uses a synthetic target constructed by the authors; no production codebase is targeted and no real-world vulnerability is disclosed in this paper. Experiment~I produced actionable harmful content from one target model; none of it is reproduced in this paper. So that the manual verification behind EHR can be audited, the anonymized supplementary material provided for review contains the raw per-attack logs, including target-model responses. The public release will separate the harness, agent roles, attack strategies, verification rubric, and per-attack scores, which will be openly licensed, from the response text of attacks verified as harmful, which will be available to researchers only on request. Experiment~I attacks were conducted via standard API endpoints under each provider's terms of service. The staged generalization study referenced in Section~\ref{sec:alt} follows responsible-disclosure protocols: pre-patch reproduction of already-fixed vulnerabilities first, then any novel-target work coordinated with affected vendors before publication. The primary near-term users we anticipate are AI labs, security teams, and procurement evaluators who need an open testbed for adversarial-robustness measurement and currently have no viable independent alternative.

\emph{Why open rather than gated.} We considered a gated or staged release of the framework (vetting requesters, or releasing capability in stages) and decided against it for this artifact, though not without reservation. The one exception is the response text of attacks verified as harmful, which we withhold from the open release as described above: it adds nothing to the testbed's defensive value and is the harmful material itself. The case for gating is the elicitation-gap argument of Section~\ref{sec:elicitation}, and we take it seriously. Three considerations decided us. First, the offensive increment is low and the capability class is already diffused: the components are coordination, role differentiation, and evolutionary search over a small open-weights model, all reconstructable from this paper alone, and in the export-control episode the developer reported that other public models found the same bugs without a bypass~\cite{anthropic2026suspend}. Second, a gate would reproduce the exact dependency the paper criticizes, making independent evaluation of frontier claims contingent on the same institutions whose claims are under test, and it would fall hardest on the under-resourced defenders, regulators, and procurement teams who most need an open testbed, while the resourced adversaries that offensive-AI policy targets are the least gate-constrained. Third, what a gate buys is time, and the defensive value of an open testbed is also time-sensitive. %We regard this as the right call for a 1.2B-scale evaluation harness, and not a general license: a framework that measurably lowered the barrier to new, high-severity exploitation would warrant the opposite call, and we would make it.

\emph{Anonymized release.} An anonymized copy of the framework, the raw logs, and the audit scripts is provided as supplementary material; the public repository will be released upon publication on the terms described above.

\emph{Reproducibility.} The supplementary material ships the raw per-attack logs and result files for both experiments together with audit scripts. The scripts recompute from the stored data and make no model calls, so their output is deterministic. For Experiment~I, the script reproduces exactly the attack counts (225 per target), the technical success rates (60.9\% and 40.0\%), the perfect-score counts (18 and 0), and the per-generation evolution figures of Table~\ref{tab:evolution} (the early-window and generations-7--14 success rates and mean scores, computed directly from the released logs). The Effective Harm Rate, the critical-severity counts, and the refusal rate come from the manual verification pass of Section~\ref{sec:exp1}. We report them as author-verified quantities; the released scripts do not recompute them. The full target responses are included in the released attack libraries, so these judgments can be checked independently against the same data, and we additionally release a script that grades every response with the published StrongREJECT rubric~\cite{souly2024strongreject} for a reproducible, harm-grounded score alongside the verified rate. Keeping pipeline output separate from author judgment in this way is the discipline Section~\ref{sec:assessment} argues evaluations should adopt. For Experiment~II, a second script recomputes the assisted recall at each swarm size (6/9, 8/9, and the combined 9/9 as the union of the source and fuzzing phases) and the autonomous crash- and citation-verified recall (0/9 and 2/9), all from the stored result files.

\section*{Use of Generative AI}

The authors used a generative AI assistant for copy-editing and LaTeX formatting during manuscript preparation. All research design, experiments, analysis, and argumentation are the authors' own, and the authors reviewed and take responsibility for all content, including all AI-assisted edits.

\bibliography{references}

\appendix

\section{Extended related work: LLM-assisted vulnerability discovery}
\label{app:related}
Prior work on using LLMs for vulnerability discovery can be organized along two axes: the realism of the target (synthetic benchmarks vs.\ production code) and the autonomy of the model (static evaluation vs.\ tool-augmented agentic operation).

\paragraph{Benchmark-based evaluation.} Evaluation suites such as CyberSecEval~2~\cite{bhatt2024cyberseceval2} present models with self-contained test cases and measure the model's ability to identify vulnerabilities and, in some configurations, produce working exploits. Useful for controlled comparison across model families, these benchmarks operate on isolated snippets and cannot capture the cross-file reasoning required for real software.

\paragraph{Agentic exploit discovery.} Industry agentic frameworks such as Google's Big Sleep (formerly Project Naptime)~\cite{bigsleep2024} pair a frontier LLM with specialized tools (e.g., code browser, sandbox, debugger) and have reported previously unknown memory-safety vulnerabilities in widely deployed software. The contribution is architectural: the LLM operates as a reasoning controller planning tool invocations over multi-step exploit chains. Both the reasoning budget and the diversity of strategies remain bounded by a single frontier model instance.

\paragraph{LLM-augmented fuzzing.} A third approach retains conventional fuzzing infrastructure but replaces or supplements its input-generation stage with an LLM. Fuzz4All~\cite{xia2024fuzz4all} uses LLMs to produce syntactically rich seed inputs across six languages and has reported many bugs in widely used systems. The LLM acts as a component, a structured generator rather than an autonomous agent, and the search strategy remains the fuzzer's.

\paragraph{Harness synthesis.} A complementary line treats the multi-agent harness itself as a search variable. AgentFlow~\cite{liu2026agentflow} automates harness search via an outer-loop optimizer reading coverage maps, sanitizer reports, and action traces, at a scale (a 192-GPU cluster for seven days on the browser target) far above ours. The two settings bracket the problem from opposite ends: if harness design can rival model choice at frontier scale on production targets, its role may be at least as large at commodity scale, where the model contributes less individual reasoning capacity. Which is a hypothesis our fixed-model design cannot test.

\paragraph{Common architectural assumption.} With the partial exception of AgentFlow's harness-search loop, these approaches share reliance on a single frontier-scale model and a \emph{monolithic topology}: one model holds the full reasoning trajectory in its context. We replace the single frontier model with a \emph{swarm} of \emph{smaller} models. There is an analogy to single-trajectory optimisation versus population-based search: individual agents are weaker, but the ensemble gains diversity of strategy, natural parallelism, and broader coverage. In Experiment~I, frontier-scale reasoning was not required to compromise GPT-4o's output safety when the search was conducted by a swarm; Experiment~II shows the complementary limit, where the swarm's own reasoning is insufficient and the scaffold carries the result.

\section{Experiment I: per-generation evolution}
\label{app:evolution}

Table~\ref{tab:evolution} reports the per-generation evolution of attack performance against each target, underlying the success-rate trajectory summarized in Section~\ref{sec:exp1}.

\begin{table}[t]
\centering
\caption{Attack performance at the two ends of the 15-generation run (generations numbered 0--14): the early window is generations 0--1 and the late window is generations 7--14. Success is the share of attacks with judge score $\geq 0.6$; Score is the mean judge score. These two windows bracket the learning curve and are \emph{not} a partition of the run--generations 2--6 fall in neither--so the overall Technical Success Rate in Table~\ref{tab:exp1} (60.9\%/40.0\%, computed over all 225 attacks across all 15 generations) is higher than either window and is not the mean of the two. The gain against GPT-4o reflects a genuine capability increase; the larger proportional gain against Claude Sonnet~4 starts from a near-zero baseline and does not translate into harmful content (EHR remains 0\%, see Table~\ref{tab:exp1}).}
\label{tab:evolution}
\begin{tabular}{lcccc}
\toprule
 & \multicolumn{2}{c}{\textbf{GPT-4o}} & \multicolumn{2}{c}{\textbf{Sonnet~4}} \\
\cmidrule(lr){2-3} \cmidrule(lr){4-5}
\textbf{Phase} & Succ. & Score & Succ. & Score \\
\midrule
Early gens (0--1) & 33.3\% & 0.313 & 6.7\% & 0.057 \\
Late gens (7--14) & 65.8\% & 0.610 & 46.7\% & 0.429 \\
\midrule
\textbf{Improvement} & \textbf{+97\%} & \textbf{+95\%} & +600\% & +657\% \\
\bottomrule
\end{tabular}
\end{table}

\section{Experiment II: scaffold iteration and scaling}
\label{app:iterations}

The scaffold was developed across five configurations against the same nine planted bugs, with the model held fixed at LFM2.5-1.2B-Thinking, so every recall gain is attributable to the scaffold rather than the model. Assisted recall scaled with swarm size: 6/9 at 3~agents~$\times$~3~generations, 8/9 at 5$\times$3, and 9/9 at 7$\times$3. Because the scaffold was tuned against these specific bugs, the assisted 9/9 is close to a fixed point of its own design; the quantity the experiment reports is the distance between assisted and autonomous recall at fixed model and target (Section~\ref{sec:exp2}), not the 9/9 itself. The last bug recovered, integer overflow (CWE-190), required a scaffold-side search for \texttt{uint32\_t} narrowing patterns and a pattern-to-CWE normalization step rather than any improvement in the model's reasoning; the per-CWE modality breakdown is in Appendix~\ref{app:modality}. The evaluation-integrity correction noted in Section~\ref{sec:attribution} belongs to this history: an earlier extractor matched \texttt{CWE-XXX} strings in the planted source and so read ground-truth labels out of code comments, inflating recall to a spurious 100\%; after the labels were stripped and the extractor restricted to regex hits, role-specialized search, and source-verified citations, recall at 5$\times$3 was 8/9. The full per-version protocol is in the supplementary material.

\section{Experiment II: autonomous configuration protocol}
\label{app:autonomous}

The assisted results measure what the full pipeline produces, not what the 1.2B model contributes alone. In the autonomous configuration the LLM must generate its own fuzz inputs; the hand-crafted seed corpus, the regex detector (for recall attribution), and post-hoc normalization are disabled. Findings are credited only via (1) \emph{citation-verified} recall, where the LLM's cited source line contains a distinctive signature of a planted bug (signatures held in a ground-truth manifest used only at scoring time, never passed to the LLM), and (2) \emph{crash-verified} recall, where an LLM-generated fuzz input produces an actual ASan crash. At 7~agents~$\times$~3~generations the model produced 6 grounded citations (2 matching ground-truth signatures), of which 1 carried the correct CWE label, and 0 fuzz inputs that triggered a crash.

\section{Experiment II: per-CWE detection modality}
\label{app:modality}

Table~\ref{tab:modality} gives the detection modality for each planted CWE in the assisted pipeline, underlying the attribution claim in Section~\ref{sec:exp2}: most recoveries come from the hand-crafted fuzz corpus and regex-matched seeds, and the model's one credited contribution is redundant with role-specialized search.

\begin{table*}[t]
\centering
\caption{Detection modality for each planted CWE in the assisted pipeline at 7~agents~$\times$~3~generations (v5). ``Source'' denotes Phase~1 (regex, role-specialized search, or grounded LLM citation with post-hoc normalization); ``fuzz'' denotes Phase~2 (ASan against the seed corpus and mutations). Several CWEs are recovered by both; CWE-798 is invisible to dynamic analysis and recovered only by source inspection; several others only by fuzzing. No single modality reaches full recall, and the LLM's one credited contribution (CWE-798) is redundant with role-specialized search.}
\label{tab:modality}
\begin{tabular}{lll}
\toprule
\textbf{CWE} & \textbf{Type} & \textbf{Found by} \\
\midrule
CWE-121 & Stack buffer overflow & source (regex) + fuzz \\
CWE-122 & Heap buffer overflow & fuzz \\
CWE-134 & Format string & fuzz (stdout heuristic) \\
CWE-190 & Integer overflow & source (arithmetic role, v5)$^{\dagger}$ \\
CWE-416 & Use-after-free & fuzz \\
CWE-415 & Double free & fuzz \\
CWE-476 & NULL pointer deref & fuzz \\
CWE-78 & Command injection & source (regex) + fuzz \\
CWE-798 & Hardcoded credentials & source (search) + LLM cite + norm. \\
\bottomrule
\end{tabular}
\begin{flushleft}
\small $^\dagger$At runtime CWE-190 manifests as a heap overflow (CWE-122); fuzzing detects the crash but attributes it to the wrong CWE. Source-level attribution comes from the v5 arithmetic role's search for \texttt{uint32\_t} multiplication patterns, mapped to CWE-190 by the pattern-to-CWE table.
\end{flushleft}
\end{table*}

\section{Extended limitations}
\label{app:limitations}
The LLM-grounded code analysis varies highly across runs: the same configuration may find 0 or 4 ground-truth CWEs via grounded citations in different runs. The model frequently cites real vulnerable code but assigns the wrong CWE class, producing classification false positives useful as suspicious-code flags but not as precise identifications. At 1.2B, peak grounded-citation recall is 4/9 even with an increased query budget, and bugs requiring cross-file arithmetic reasoning (CWE-190) or purely runtime phenomena (CWE-122 via fuzzing) do not lift with more source-analysis budget. Phase~2 fuzzing in the assisted configuration relied primarily on hand-crafted seed inputs; the autonomous configuration's 0/9 crash-verified result shows that the 1.2B model alone cannot generate replacements at this scale.

\end{document}